\documentclass[referee]{raa}           
\usepackage{graphicx,times}
\usepackage{natbib}
\usepackage{amssymb,amsmath}
\bibpunct{(}{)}{;}{a}{}{,}

\usepackage[font=small,labelfont=bf,labelsep=space]{caption}
\usepackage[a4paper=true,dvipdfm=true,pagebackref=true]{hyperref}
\hypersetup{pdftitle = The title of my PDF, pdfauthor = My name, pdfsubject= The subject, pdfkeywords = keyword1 keyword2 keyword3} 
\hypersetup{colorlinks = true, linkcolor = green, anchorcolor = red, citecolor = blue, filecolor = red, pagecolor = red, urlcolor = red}

\newcommand{\kms}{\mbox{$\mathrm{km\,s^{-1}}$}}
\newcommand{\Ha}{\mbox{${\mathrm H\alpha}$}}
\newcommand{\Ion}[2]{#1{\,\scriptsize #2}}

\newcommand{\Lin}[3]{\Ion{$[$#1}{#2}$]$\,$\lambda$\,#3}
\newcommand{\Li}[3]{\Ion{$[$#1}{#2}$]$\,#3}
\newcommand{\Lines}[3]{\Ion{$[$#1}{#2}$]$\,$\lambda\lambda$\,#3}
\newcommand{\teff}{\mbox{$T_{\rm eff}$}}
\newcommand{\logg}{\mbox{$\log g$}}
\newcommand{\feh}{\mbox{$\rm{[Fe/H]}$}}

\begin{document}

   \title{Mapping the emission line strengths and kinematics of Supernova Remnant S147 with extensive LAMOST spectroscopic observations}

 \volnopage{ {\bf 2018} Vol.\ {\bf X} No. {\bf XX}, 000--000}
   \setcounter{page}{1}

   \author{Juan-Juan Ren\inst{1,3}\thanks{LAMOST Fellow}, Xiao-Wei Liu\inst{1,2,5}, Bing-Qiu Chen\inst{1,5}\footnotemark[1], Mao-Sheng Xiang\inst{3}\footnotemark[1], Hai-Bo Yuan\inst{4}, Yang Huang\inst{1,5}\footnotemark[1], Hua-Wei Zhang\inst{1}, Chun Wang\inst{1}, Zhi-Jia Tian\inst{1}\footnotemark[1], Gao-Chao Liu\inst{1,6}, Hong Wu\inst{3}}

   \institute{ Department of Astronomy, Peking University, Beijing 100871, P.\,R.\,China; {\it jjren@pku.edu.cn, x.liu@pku.edu.cn}\\ 
       \and
            Kavli Institute for Astronomy and Astrophysics, Peking University, Beijing 100871, P.\,R.\,China;\\
       \and
            Key Laboratory of Optical Astronomy, National Astronomy Observatories, Chinese Academy of Sciences, Beijing 100012, P.\,R.\,China;\\
       \and
            Department of Astronomy, Beijing Normal University, Beijing 100875, P.\,R.\,China;\\
       \and
            South-Western Institute for Astronomy Research, Yunnan University, Kunming, Yunnan 650091, P.\,R.\,China;\\
       \and
			College of Science, China Three Gorges University, Yichang 443002, P.\,R.\,China\\       
}


\abstract{We present extensive spectroscopic observations of supernova remnant (SNR) S147 collected with the Large sky Area Multi-Object fiber Spectroscopic Telescope (LAMOST). The spectra were carefully sky-subtracted taking into account the complex filamentary structure of S147. We have utilized all available LAMOST spectra toward S147, including sky and stellar spectra. By measuring the prominent optical emission lines including \Ha, \Lin{N}{II}{6584}, and \Lines{S}{II}{6717,\,6731}, we present maps of radial velocity and line intensity ratio covering the whole nebula of S147 with unprecedented detail. The maps spatially correlated well with the complex filamentary structure of S147. For the central 2$^\circ$ of S147, the radial velocity varies from $-$100 to 100\,\kms\ and peaks between $\sim$\,0 and 10\,\kms . The intensity ratios of \Ha/\Lines{S}{II}{6717,\,6731}, \Lin{S}{II}{6717}/$\lambda$\,6731 and \Ha/\Lin{N}{II}{6584} peak at about 0.77, 1.35 and 1.48, respectively, with a scatter of 0.17, 0.19 and 0.37, respectively. The intensity ratios are consistent with the literature values. However, the range of variations of line intensity ratios estimated here and representative of the whole nebula, are larger than previously estimated.
}

   \authorrunning{J.-J. Ren et al. }            
   \titlerunning{SNR S147}  
   \maketitle

%
\section{Introduction}           
\label{sect:intro}

Supernova remnants (SNRs) are interesting objects to study for a number of reasons. They provide insights into the mechanisms of supernova explosions and are possible sources of the Galactic cosmic rays. Moreover, SNRs probe the immediate surroundings of supernovae, shaped by their progenitors. Currently 294 Galactic SNRs \citep{Green2014BASI...42...47G} have been discovered. A great majority ($\sim$\,79\%) of the known Galactic SNRs are believed to be relatively old (with ages greater than 10$^3$ years and radii larger than $\sim$\,5\,pc) and evolved objects in either the adiabatic or the early radiative stages of their evolutionary developement \citep{Woltjer1972ARA&A..10..129W}, and generally show characteristic shell structures \citep{Fesen1985ApJ...292...29F}. About 30\% of Galactic SNRs are known to have optical emission associated with their nonthermal radio emission. For old remnants, the optical emission arises from the cooling of shocked interstellar cloud material following the passage of the remnant's blast waves as they expand outward into the ambient medium. Optical spectra of filaments of evolved SNRs show strong emission including that of \Ha, \Lines{O}{II}{3726,3729}, \Lines{O}{III}{4959,5007}, \Lines{N}{II}{6548,6584} and \Lines{S}{II}{6717,6731}. The line intensity ratios have important physical implications, such as \Ha/\Lines{S}{II}{6717,6731} ratio can be used to distinguish shocked nebulae from photoionized nebulae; the \Lin{S}{II}{6717}/\Lin{S}{II}{6731} ratio is electron density sensitive; the \Ha/\Lin{N}{II}{6584} ratio is a widely used tool for investigating nitrogen-to-hydrogen abundance variations among SNRs. Thus it important to obtain complete sampling of the optical line emissions in SNRs. Although several evolved remnants have been well studied morphologically in the optical, only a limited amount of spectroscopic data are available for those faint optical remnants, especially for those with large angular sizes including SNR S147. 

\begin{figure}
    \centering
	\includegraphics[angle=0,width=0.5\textwidth]{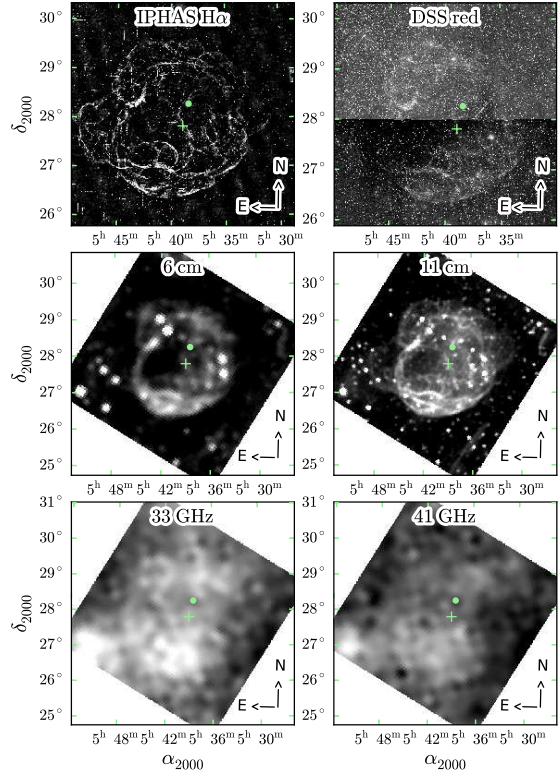}
	\caption{Multi-band images of SNR S147. The upper panel shows optical images in \Ha\ (left) from the IPHAS survey \citep{Drew2005MNRAS.362..753D} and in $R$-band from the DSS (right), respectively. The middle panel shows respectively radio images at 11\,cm (2639\,MHz; left) and 6\,cm (4800\,MHz; right) obtained with the Effelsberg 100-m and the Urumqi 25-m telescopes respectively \citep{Xiao2008A&A...482..783X}. The bottom panel shows respectively high-frequency radio maps at 31\,GHz and 44\,GHz obtained with the WMAP \citep{Xiao2008A&A...482..783X}. The light green cross and dot show respectively the geometrical centre of S147 and the position of pulsar J0538+2817, probably associated with S147. }
	\label{fig:S147images}
\end{figure}

S147 (also named G180.0$-$1.7) is an optically faint, highly filamentary shell-type SNR in the direction of Galactic-anticentre. It was first identified as a SNR candidate by \citet{Minkowski1958RvMP...30.1048M}. S147 is now believed to be one of the most evolved SNRs (with an estimated age $\sim$\,10$^5$ yr) in the Galaxy. Except for the spurs associated with a SNR, S147 has a shell-like structure of diameter $\sim$\,200$'$. In the optical bands, the shell structure is dominated by filamentary H$\alpha$ emission. The emission is bright at the northern and southern edges but concentrates mainly in the southern parts \citep{Dincel2015MNRAS.448.3196D}. Despite of its large age, it conserves well its spherical symmetry except for the blowout regions in the east and west. The optical and radio shell morphologies are well defined \citep{Minkowski1958RvMP...30.1048M,Fuerst1986A&A...163..185F}, and coincide with each other. Fig.\,\ref{fig:S147images} shows images of S147 in different bands. Although no X-ray emission has been detected by EXOSAT \citep{Sauvageot1990A&A...227..183S}, \cite{Sun1996rftu.proc..195S} report unambiguous X-ray emission from S147 based on data from the ROSAT All Sky Survey \citep[RASS,][]{Voges1999A&A...349..389V}. \cite{Chen2017MNRAS.472.3924C} claim that the X-ray emission and the dust extinction is anti-correlated. A spatially extended gamma-ray source detected in the energy range 0.2\,--\,10\,GeV is found to coincide with S147. The gamma-ray emission exhibits a possible spatial correlation with the prominent \Ha\ filaments of S147 \citep{Katsuta2012ApJ...752..135K}.

A compact object, the radio pulsar J0538+2817 \citep{Anderson1996ApJ...468L..55A}, located 40$'$ west of the centre of S147, is believed to be associated with S147. An early type (B0.5V) runaway star, HD 37424, has also been found inside S147, believed to be the pre-supernova binary companion to the progenitor of the pulsar and the SNR \citep {Dincel2015MNRAS.448.3196D}. The distance estimated of S147 in the literature spans a wide range, 0.6\,--\,1.9\,kpc, depending on the methods used \citep{Dincel2015MNRAS.448.3196D}. The surface brightness and distance ($\Sigma$--D) relation typically yields a distance of $\sim$\,1\,kpc from several studies \citep{Clark1976MNRAS.174..267C, Kundu1980A&A....92..225K, Guseinov2003A&AT...22..273G}. The interstellar reddening toward S147 suggests a smaller distance of 0.8\,kpc \citep{Fesen1985ApJ...292...29F}. On the other hand, distances derived from the parallax of the pulsar are significantly larger, ranging between 1.3\,--\,1.5\,kpc \citep{Chatterjee2009ApJ...698..250C, Ng2007ApJ...654..487N}. A recent study yields distance $d = 1.333^{+0.103}_{-0.112}$\,kpc and extinction $A_V = 1.28 \pm 0.06$\,mag \citep{Dincel2015MNRAS.448.3196D}. Finally, from a dust extinction analysis based on data from the Xuyi Schmidt Telescope Photometric Survey of the Galactic Anti-centre \citep[XSTPS-GAC;][]{Liu2014IAUS..298..310L,Zhang2014RAA....14..456Z}, \citet{Chen2017MNRAS.472.3924C} obtain a new measurement of distance to S147, $d = 1.22 \pm 0.21$\,kpc. The expansion velocity estimated for the S147 shell ranges between 80\,--\,120\,\kms\ \citep{Kirshner1979ApJ...229..147K,Phillips1981MNRAS.195..485P}.

Previous optical spectrophotometric observations of S147 \citep{Kirshner1979ApJ...229..147K, Fesen1985ApJ...292...29F} obtain only dozens of spectra, covering a small portion of the extremely large and faint S147 SNR. The observations focused on several bright filaments. S147 consists of numerous filaments embedded in diffuse emission. The two components (i.e., bright filaments and diffuse emission) have quite different physical properties. It is thus important to sample fully the whole extent of S147 in order to obtain a comprehensive understanding of an evolved SNR such as S147.
 
\begin{figure}
    \centering
	\includegraphics[angle=270,width=150mm]{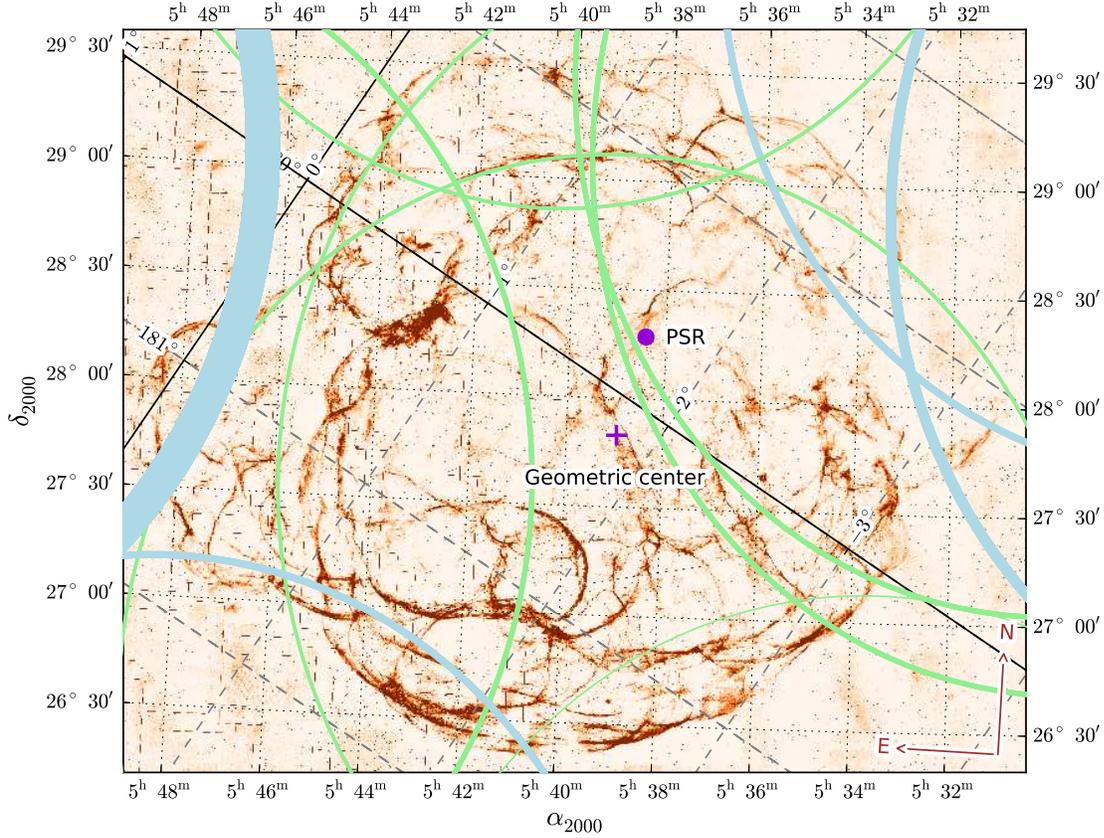}
    \caption{LAMOST observations in the direction of S147, overlaid on its IPHAS \Ha\ image \citep{Drew2005MNRAS.362..753D}. Light green circles represent 16 LAMOST plates centred within 3.55$^\circ$ of the geometric centre of S147, while light blue circles show 34 LAMOST plates centred at larger distances. Both the light green and blue circles have a radius of 2.5$^\circ$, the radius of the field of view of LAMOST. The thickness of the circles represents the number of repeated observations. The violet cross shows the geometric  centre of S147, while the violet dot denotes pulsar J0538+2817. The Figure is shown in the equatorial coordinate system, with grids delineated by dotted lines. The grey dashed lines show grids in the Galactic coordinate system, and the two orthogonal black solid lines denotes $l$\,=\,180$^\circ$ and $b$\,=\,0$^\circ$, respectively.}
    \label{fig:lamost_plates}
\end{figure}

The Large sky Area Multi-Object fiber Spectroscopic Telescope (LAMOST) is a quasi-meridian reflecting Schmidt telescope of $\sim$\,4-m effective aperture, with a field of view of 5$^\circ$ diameter \citep{Cui2012RAA....12.1197C}. It is located in Xinglong station of National Astronomical Observatories of Chinese Academy of Sciences. Being a dedicated survey telescope, LAMOST takes spectra with 4\,000 fibers, targeting interested celestial objects including sky background and calibration sources in one single exposure with 16 fiber-fed spectrographs, each accommodates 250 fibers. The spectra cover the entire optical wavelength range ($\simeq$\,3700\,--\,9000\AA) at a resolving power $R$\,$\sim$\,1800. Since 2012 September, LAMOST has been carrying out a five-year Regular Surveys. Before that there was a one-year Pilot Surveys preceded by a two-year commissioning phase. The LAMOST Regular Surveys consist of two components \citep{Zhao2012RAA....12..723Z}: the LAMOST Extra-Galactic Survey of galaxies (LEGAS) that aims at studying the large scale structure of the universe, and the LAMOST Experiment for Galactic Understanding and Exploration (LEGUE) that aims at obtaining millions of stellar spectra in order to study the structure and evolution of the Milky Way galaxy \citep{Deng2012RAA....12..735D}. 

The LEGUE has three sub-components: the spheroid, the anticentre and the disc surveys. The LAMOST Spectroscopic Survey of the Galactic Anticentre \citep[LSS-GAC;][]{Liu2014IAUS..298..310L,Yuan2015MNRAS.448..855Y,Xiang2017MNRAS.467.1890X} aims to survey a significant volume of the Galactic thin/thick discs and halo for a contiguous sky area of over 3400\,deg$^2$ centred on the Galactic anticentre ($|b| \leq 30^\circ$, 150$^\circ \leq l \leq$ 210$^\circ$), down to a limiting magnitude of $r$\,$\sim$\,17.8\,mag (to 18.5\,mag for limited fields). S147 is located within the LSS-GAC footprint. A large number of foreground and background stars toward S147 have been targeted by LAMOST accidentally, some by several times. S147 has a large diameter and contains many filaments. It would be extremely time consuming if one attempts to obtain spectra covering its full spatial extent using regular telescopes. The wide field of view, multi-fiber survey telescope LAMOST thus serves as a perfect tool for mapping the details of this extremely extended, evolved SNR. It provides us a great opportunity to study global properties of S147 with unprecedented detail for the first time. 

In this paper, we focus on the analysis of the LAMOST optical spectra of S147. In particular, we concentrate on the kinematic properties revealed by strong emission lines including \Ha, \Lines{N}{II}{6548,6584} and \Lines{S}{II}{6717,6731}. In Section 2 we describe the observation and data reduction. In Section 3 we present parameter determinations and result analysis. Section 4 summarize our findings.

\section{Observation and Data Reduction}

\subsection{Observation}

The LAMOST Data Release 2 (DR2) includes data from 1\,934 plates collected by June 2014, amongst them 401 plates are from the Pilot Surveys (from October 2011 to June 2012), 808 plates from the first year Regular Surveys (from September 2012 to June 2013), and 725 plates from the second year Regular Surveys (from September 2013 to June 2014). Totally, the DR2 general catalogue contains 4\,132\,782 spectra, including 3\,779\,674 stellar, 37\,665 of galaxies, 8\,633 of QSOs, and 306\,810 of unknown objects. The LAMOST DR2 data was released internally on December 2014 and publicly on June 2016 \footnote{http://dr2.lamost.org/}.

Being a large-size SNR in the direction of Galactic anticentre, many foreground and background stars in the vicinity region of S147 have been targeted, some by several times by LAMOST since 2011. There are 50 LAMOST plates overlapping with (or very close to) S147. Fig.\,\ref{fig:lamost_plates} plots the positions of these 50 plates included in the LAMOST DR2, overlaid with an \Ha\ image from the IPHAS survey \citep{Drew2005MNRAS.362..753D}. 

Amongst the 50 LAMOST plates, 16 have a centre of field of view within 3.55$^\circ$ of the geometric centre of S147 (10 B plates, and 6 M plates \footnote{LAMOST plates are grouped into four categories based on the magnitude ranges of the targets: Very Bright (VB; $r$\,$\leq$\,14\,mag), Bright (B; 14\,$<$\,$r$\,$<$\,16.8\,mag), Medium-bright (M; 16.8\,$\leq$\,$r$\,$\leq$\,17.8\,mag), and Faint (F; $r$\,$>$\,17.8\,mag) \citep{Luo2015RAA....15.1095L,Liu2014IAUS..298..310L,Yuan2015MNRAS.448..855Y}.}). Of those 16 plates, 9 are from the Pilot Surveys, 7 from the first year Regular Surveys. The other 34 plates have centres of field of view between 3.55$^\circ$ and 4.54$^\circ$ from the geometric centre of S147 (5 VB plates, 26 B plates, 16 M plates and 3 F plates). Amongst those 34 plates, 28 are from the Pilot Surveys, 18 from the first year Regular Surveys, and 4 from the second year Regular Surveys.

\subsection{Data Reduction}

\subsubsection{Overview}

LAMOST raw spectra are processed with the LAMOST two-dimensional (2D) pipeline \citep{Luo2012RAA....12.1243L, Luo2015RAA....15.1095L}, including dark and bias subtractions, cosmic ray removal, one-dimensional (1D) spectral extraction, flat-field correction, wavelength calibration, sky subtraction, merging sub-exposures, and finally, splicing the sub-spectra from the blue- and red-arm of the spectrograph, respectively. \citet{Xiang2015MNRAS.448...90X} develop an iterative algorithm for the flux calibration in order to overcome the effect of often heavy interstellar reddening that hinders the selection of suitable standard stars (F stars). For a given spectrograph, the spectra are first flux-calibrated using the nominal spectral response curve (SRC) and the initial stellar atmospheric parameters are derived with the LAMOST Stellar Parameter Pipeline at Peking University \citep[LSP3;][]{Xiang2015MNRAS.448..822X}. Then based on those initial estimates of stellar parameters, stars of effective temperature \teff\ between 5750 and 6750\,K (mostly of F spectral type) are selected as the flux-calibration standard stars and used to deduce an updated SRC after corrected for the interstellar reddening, derived by comparing the observed and synthetic photometric colours, assuming the \citet{Fitzpatrick1999PASP..111...63F} extinction law for a total to selective extinction ratio $R_V$\,=\,3.1. The new SRC is then used to flux-calibrate the spectra and the stellar parameters are updated accordingly from the recalibrated spectra. The above process is repeated until a convergence is achieved. Usually more than four standard stars per spectrograph can be identified and selected for the majority of the LSS-GAC plates. In cases where not enough standard stars can be selected for a given spectrograph (plate), the SRC(s) deduced from other plates, usually collected on the same night, are used to do the calibration. A comparison of spectra obtained at different epochs for duplicate targets indicates that the relative flux calibration has achieved an accuracy of about 10\,per\,cent for spectra of a signal to noise (SN) ratio  per pixel higher than 30 at 4650\,\AA\  \citep{Xiang2015MNRAS.448...90X}. 

\begin{figure}
    \centering
	\includegraphics[angle=0,width=85mm]{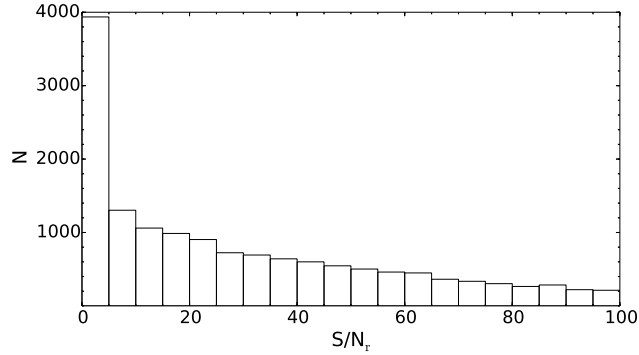}
    \caption{Histogram distributions of spectral SN ratio in red (right panel) parts of the spectra, for the 19\,736 LAMOST spectra within 2$^\circ$ of S147.}
    \label{fig:sn_hist}
\end{figure}

\begin{figure}
    \centering
	\includegraphics[angle=0,width=85mm]{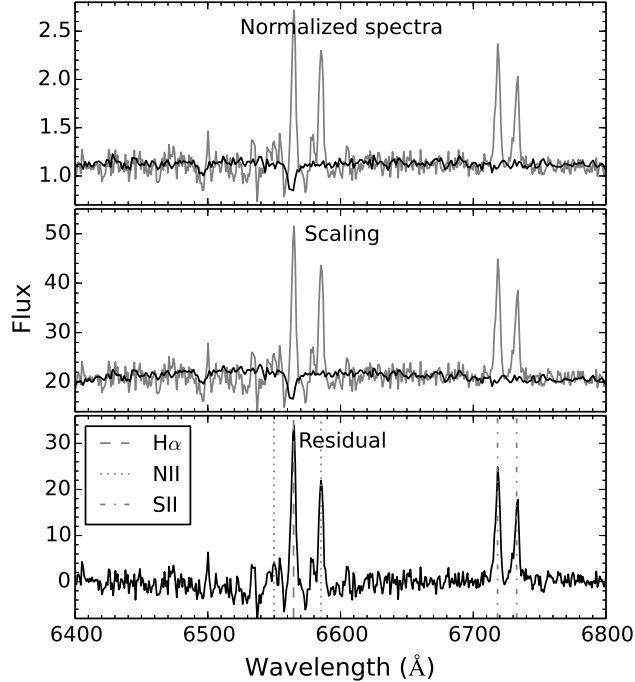}
    \caption{Example of stellar spectrum subtraction. The first row shows the normalized spectrum of a target star of LAMOST spectral ID ``20111126-GAC$\_$082N27$\_$M1-3-85" (grey line), and that of the paired-star of ID ``20120203-B5596104-6-55" (black line). The second row shows the same two spectra after the continuum level of the paired-star spectrum has been scaled to match that of the target star. The residuals of subtraction (i.e. the spectrum of S147 filament) are shown in the third row.}
    \label{fig:substar}
\end{figure}
\subsubsection{Data toward S147}

In total, there are 19\,736 spectra (including those from the sky fibers) within 2$^\circ$ of the geometric centre of S147, of which 4\,667 spectra (4\,349 unique stars) have available LSP3 stellar atmospheric parameters. Fig.\,\ref{fig:sn_hist} shows the histogram distributions of spectral SN ratio in the red (7450\,\AA) regions of the 19\,736 spectra within 2$^\circ$ of S147. 

\subsubsection{Sky subtraction}

For each LAMOST plate, $\sim$\,10\,--\,15\,per\,cent of the fibers are assigned as object-free sky fibers pointing toward areas of blank sky for modeling the night sky background. A method of B-spline fitting \citep{Stoughton2002AJ....123..485S,Luo2015RAA....15.1095L} is applied to make a supersky for each sub-exposure of a spectrograph and then the principal component analysis (PCA) is used to recorrect the supersky. The fiber assignment strategy works well for fields at high Galactic latitudes, yielding results comparable to those of the SDSS survey. However, for sky areas near the Galactic plane, such as of S147 where the sky is plagued by diffuse and highly filamentary nebular emission, the strategy fails to build a supersky of adequate accuracy. In the current work, we develop a new method for the sky subtraction, which is described below.

Overall S147 has a quite regular and nearly spherical shape, with a diameter of $\sim$\,3$^\circ$, smaller than the field of view of LAMOST (5$^\circ$ in diameter). For the individual LAMOST plates taken in the vicinity field of S147, there are always some regions not contaminated by the emission of S147. Thus for those plates, we are able to use the sky fibers positioned at points uncontaminated by S147 emission to rebuild the supersky, and redo the sky subtraction. This procedure is carried out for the blue- and red-arm  spectra separately. The uncontaminated sky spectra from the non-S147 region are selected following the criteria: 1) At least 2$^\circ$ away from the S147 geometric centre; 2) Not contaminated by S147 emission, i.e., those with a detectable H$\alpha$ flux (larger than 2$\sigma$, for strictly cut) are also excluded. The sky spectra thus selected and satisfying the above criteria are then averaged to construct the so-called ``uncontaminated non-S147 supersky". 

LAMOST has 16 spectrographs, each feeds by 250 fibers, where 10 to 20 of them are normally assigned as sky fibers. For the individual spectrographs, the ``uncontaminated non-S147 supersky" are scaled to match each sky spectrum. The scaling procedure includes two steps, scaling by the continuum and by the telluric emission lines. The continuum scaling is carried out first, and then the line scaling. In the current work intensities of the sky emission lines $\lambda$\,6498, $\lambda$\,6533 in red-arm are used for the line scaling. We then obtain a ``new initial supersky" for each of the 16 spectrographs (abnormal sky spectra are excluded during this process). We note that continuum and line scaling works much better for a limited wavelength region than for the whole spectrum. In this paper, we focus only on selected wavelength regions of interest here: 6400\,--\,6800\,\AA\ in the red that contains \Ha , and the \Lines{N}{II}{6548,6584} and \Lines{S}{II}{6717,6731} nebular lines. The sky subtraction is only carried out for the two selected wavelength regions. Once the ``new initial supersky" has been obtained, segments of this spectrum are replaced by the ``uncontaminated non-S147 supersky" for the following wavelength ranges: 6490\,--\,6620 \AA\ and 6695\,--\,6750\,\AA\ . The supersky spectra thus obtained are then used for the final sky subtraction. 

For an uncontaminated spectrograph, sky fibers fall outside the S147 emission region and are thus free of S147 emission. The final adopted supersky differs little from the supersky yielded by the LAMOST 2D pipeline. For uncontaminated spectrographs, the reconstructed new supersky agrees well with the original LAMOST supersky, indicating that the LAMOST default supersky works well for uncontaminated sky regions. For contaminated spectrographs, the differences between our new adopted supersky and that given by the LAMOST 2D pipeline can be significant, especially for spectral regions affected by prominent nebular emissions such as \Ha\ , \Lines{N}{II}{6548,6584}, and \Lines{S}{II}{6717,6731}. Our sky subtraction procedure has a typical uncertainty of approximately 10\,per\,cent. After the sky subtraction, nebular features, especially \Ha, \Lines{N}{II}{6548,6584}, and \Lines{S}{II}{6717,6731} lines become clearly visible.

\begin{figure}
    \centering
	\includegraphics[angle=0,width=85mm]{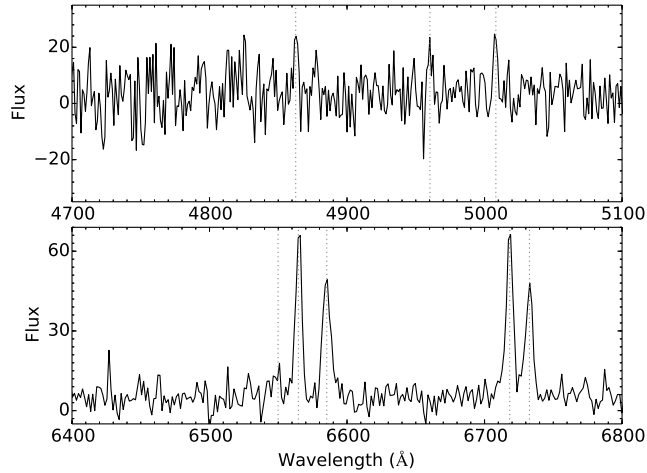}
    \caption{Example of a low SN ratio stellar spectrum of LAMOST ID ``20111024-F5909-13-28" (S/N$_\mathrm{b}$\,=\,1.0476, S/N$_\mathrm{r}$\,=\,2.8407). The top and bottom panels show the blue- and red-arm spectra, respectively.}
    \label{fig:spec-lowsnr}
\end{figure}

\begin{table*}
	\centering
	\caption{Radial velocities and intensities determined from the \Lin{N}{II}{6584} nebular line of S147. Flags a, b, and c denote results derived from sky and stellar spectra with/without stellar atmospheric parameters available from LSP3, respectively. The last column ``line" denotes the line used, i.e. \Lin{N}{II}{6584} for the example entries listed here. Note that the RV error provided here is the total error which is estimated by quadratically adding the uncertainty in the zero-point of the LAMOST wavelength calibration (10 \kms , \cite{Luo2012RAA....12.1243L}) and the error in the position of the nebular lines determined from the Gaussian fits. Full Table listing results from all spectra and all the lines analyzed, i.e., \Ha , \Lin{N}{II}{6584}, \Lin{S}{II}{6717} and \Lin{S}{II}{6731}, can be found online.}
	\label{tab:res_param}
	\setlength{\tabcolsep}{2pt}
	\begin{tabular}{lcccrrrrrrrc} 
		\hline
		 SpectrID	&	Flag	&	RA	&	Dec	&	S/N$_\mathrm{b}$	&	S/N$_\mathrm{r}$	&	${\mathrm{V_r}}$	&	Error	& Intensity & Line\\
		 &          &  &  & & & (km/s) & (km/s) & (10$^{-17}$ & \\
		 &          &  &  & & &        &       & erg/cm$^{2}$/s) & \\

		\hline
20111215-GAC$\_$082N27$\_$B1-7-145	&	b	&	05:59:44.65	&	26:17:10.8 	&	31.2	&	46.7	&	36.6	 &	12.6&	27.1  &	\Lin{N}{II}{6584}	\\
20111215-GAC$\_$082N27$\_$B1-6-109	&	b	&	05:59:43.78 	&	26:23:31.0	&	13.9	&	12.5	&	0.8	 &	12.6&	102.7 &	\Lin{N}{II}{6584}	\\
20111215-GAC$\_$082N27$\_$B1-7-134	&	b	&	05:59:30.64	&	26:07:59.3	&	12.4	&	19.9	&	21.5	 &	11.3	&	163.9 &	\Lin{N}{II}{6584}	\\
20111126-GAC$\_$082N27$\_$M1-7-140	&	a	&	05:59:25.41	&	25:59:41.9	&	2.9	&	0.9 	&	-17.2&	11.4	&	42.6	  &	\Lin{N}{II}{6584}	\\
20111215-GAC$\_$082N27$\_$B1-7-139	&	b	&	05:59:23.28	&	26:14:29.6	&	14.6	&	20.7	&	11.6	 &	11.7&	169.6 &	\Lin{N}{II}{6584}	\\
20111126-GAC$\_$082N27$\_$M1-7-94	&	a	&	05:59:03.98	&	25:59:41.9	&	8.1 &	2.8 &	-28.3&	13.3	&	33.7  &	\Lin{N}{II}{6584}	\\
20111126-GAC$\_$082N27$\_$M1-7-135	&	b	&	05:58:43.02	&	26:07:58.6	&	11.6	&	29.4	&	8.1 	 &	11.3	&	99.4	  &	\Lin{N}{II}{6584}	\\
20111215-GAC$\_$082N27$\_$B1-7-135	&	b	&	05:58:42.08 & 	26:11:06.4	&	18.1	&	35.1	&	36.2	 &	10.8&	185.9 &	\Lin{N}{II}{6584}	\\
20111215-GAC$\_$082N27$\_$B1-7-181	&	b	&	05:58:24.94	&	26:11:54.3	&	33.5	&	71.1	&	22.4	 &	15.1	&	104.8 &	\Lin{N}{II}{6584}	\\
20111215-GAC$\_$082N27$\_$B1-7-186	&	b	&	05:58:12.60	&	26:07:15.9	&	16.5	&	28.4	&	28.5	 &	11.4&	127.1 &	\Lin{N}{II}{6584}	\\
		\hline
		\end{tabular}
\end{table*}

\begin{figure}
    \centering
	\includegraphics[angle=0,width=85mm]{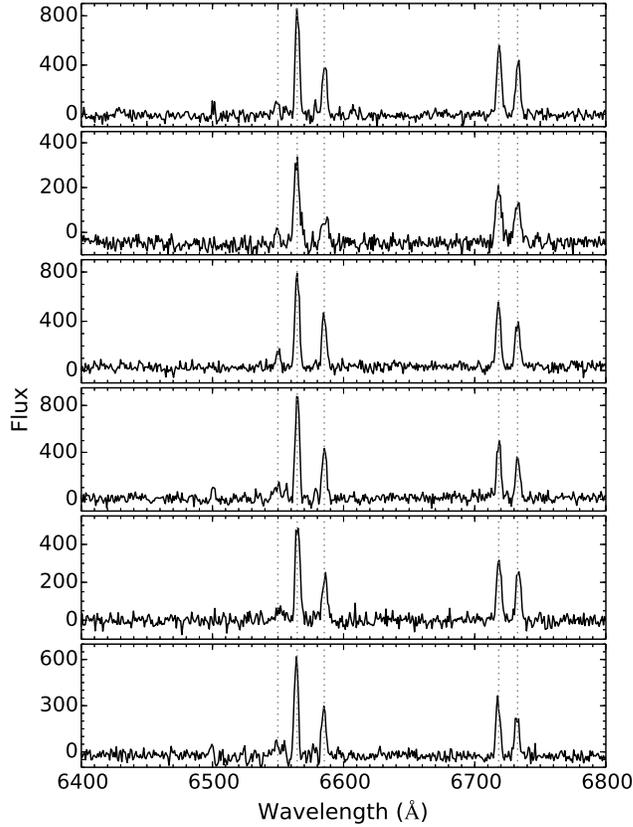}
    \caption{Example spectra in red arm for several selected filaments.}
    \label{fig:spec-filament}
\end{figure}

\begin{figure}
    \centering
	\includegraphics[angle=0,width=150mm]{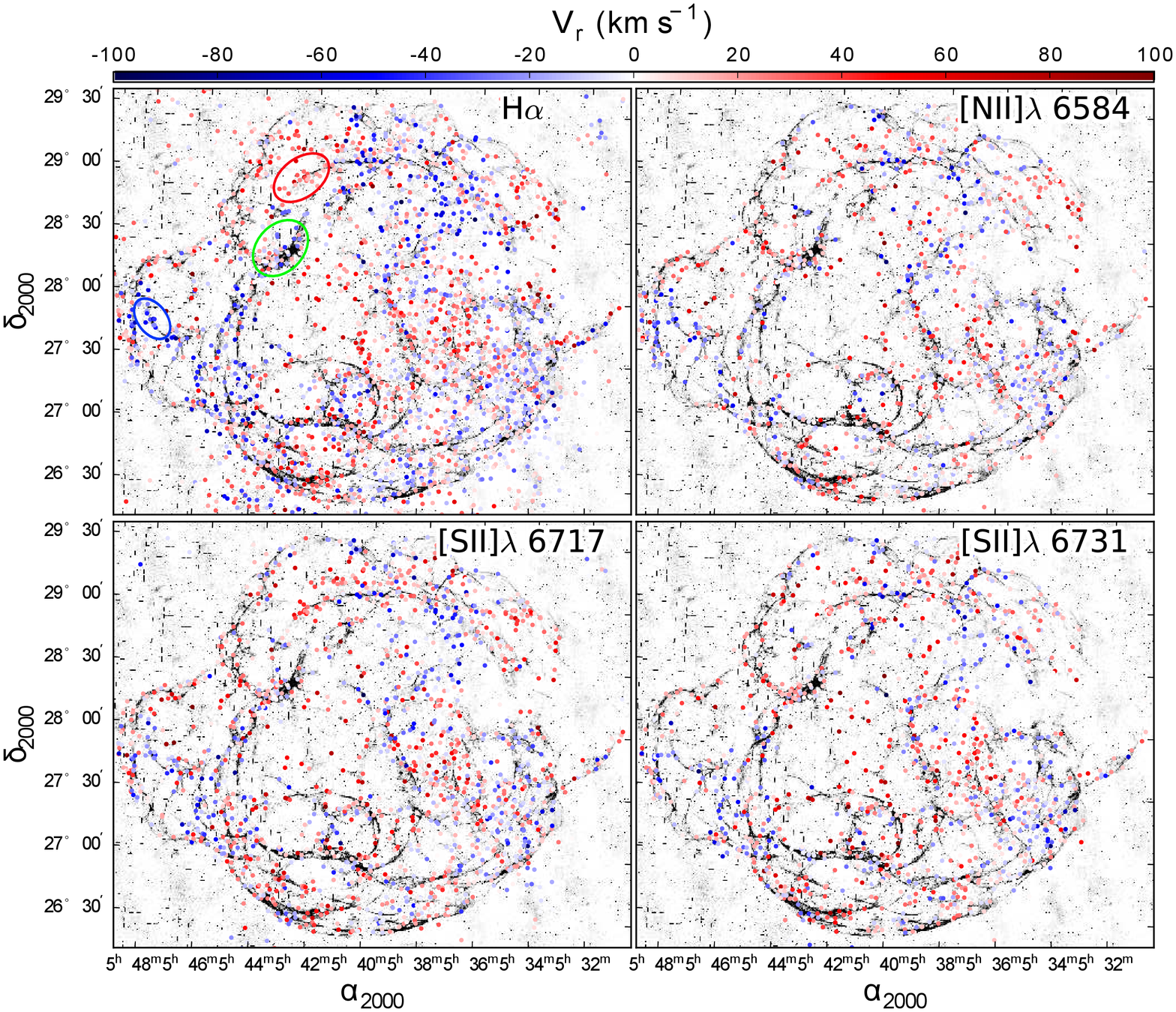}
    \caption{S147 radial velocity field as revealed by \Ha , \Lin{N}{II}{6584}, and \Lines{S}{II}{6717,\,6731} emission lines. The color bar shows the radial velocity values (\kms). Within 3$^\circ$ of the geometric centre of S147, there are in total 3344, 1282, 1707, 1310 data points in the panels from top left to bottom right, respectively. The blue and red ellipses mark example filaments moving away and towards the observer, respectively. The green ellipse shows an example of blue-/red-shifted filaments tangled together.}
    \label{fig:rv-field}
\end{figure}

\section{Parameter determinations}

To investigate the kinematic properties of S147, we determine by Gaussian fitting the radial velocities and line intensities of the prominent emission lines including \Ha , \Lines{N}{II}{6548,\,6584}, and \Lines{S}{II}{6717,\,6731}. Only nebular lines in the red-arm spectra are analyzed, given that they are much more stronger than those in the blue. Usually, the red-arm spectra have higher SN ratio than the blue-arm ones. 

There are typically $\sim$\,300 sky spectra per plate. To utilize as many spectra as possible, all spectra of foreground and background stars in the S147 vicinity field are analyzed as well. A small portion of those stellar spectra have good SN ratio, allowing stellar parameter determinations. For the remaining ones, because of the low SN ratio, no stellar parameters are available from LSP3 (as shown in Fig.\,\ref{fig:sn_hist}). Since the stellar features may affect the measurements of nebular emission lines of S147, they need to be subtracted before determining the properties of nebular emission lines. 

For high quality stellar spectra with available stellar atmospheric parameters determined with LSP3, the subtraction is done using high quality spectra of ``paired stars" that have almost the same stellar atmospheric parameters in regions uncontaminated by nebular emission \citep{Yuan2013MNRAS.430.2188Y}. The ``paired stars" are selected from the LSS-GAC Value-added Catalogues DR1 \citep{Yuan2015MNRAS.448..855Y} following the criteria: 1) Located at high Galactic latitudes ($b$\,$>$\,20$^\circ$); 2) Suffer from low extinction (ebv$\_$sp\,$<$\,0.05\,mag, where ebv$\_$sp is the $E(B-V)$ derived from the star pair method, see \citet{Yuan2015MNRAS.448..855Y} for details); 3) Have high SN ratio spectra (S/N$_\mathrm{b}$\,$>$\,20 or S/N$_\mathrm{r}$\,$>$\,20; In a few cases where no matching stars are found, this is relaxed to S/N$_\mathrm{b}$\,$>$\,10 or S/N$_\mathrm{r}$\,$>$\,10, where S/N$_\mathrm{b}$ and  S/N$_\mathrm{r}$ represent the signal to noise ratio in $b$ and $r$ band respectively); 4) Have stellar atmospheric parameters of small uncertainties (effective temperature error teff$\_$err\,$<$\,150 K, surface gravity error logg$\_$err\,$<$\,0.25\,dex, metallicity error feh$\_$err\,$<$\,0.15\,dex); 5) Have stellar atmospheric parameters that differ from the targeted stars by small amounts (specifically, the effective temperature difference $\Delta$\teff\,$<$\,150 K, surface gravity difference $\Delta$\logg\,$<$\,0.25\,dex, metallicity difference $\Delta$\feh\,$<$\,0.15 dex). For a given target star, the spectrum of a ``paired star" that satisfies all above criteria and has the smallest value of $d = \frac{\Delta\teff}{150} + \frac{\Delta\logg}{0.25} + \frac{\Delta\feh}{0.15}$ is used to subtract the stellar spectrum from the target spectrum. Fig.\,\ref{fig:substar} gives an example of stellar spectrum subtraction.

Most of those low SN ratio stellar spectra with no stellar atmospheric parameters available from LSP3 are of faint stars. The spectra are thus only marginally affected by stellar features. For those spectra, we have ignored the possible contamination of stellar features and measure directly properties of the nebular emission lines without stellar spectrum subtraction. Fig.\,\ref{fig:spec-lowsnr} shows an example of a low SN ratio stellar spectrum. Although the (stellar) continuum is quite noisy and has flux close to zero, the nebular emission lines are quite prominent. 

We determine radial velocities and intensities of \Ha\ , \Lines{N}{II}{6548,\,6584}, and \Lines{S}{II}{6717,\,6731} nebular emission lines from all available spectra in the vicinity field of S147. Part of the results are given in Table 4. The full Table can be found online. Fig.\,\ref{fig:spec-filament} shows example red-arm spectra for several selected filaments.

\begin{figure}
    \centering
	\includegraphics[angle=0,width=85mm]{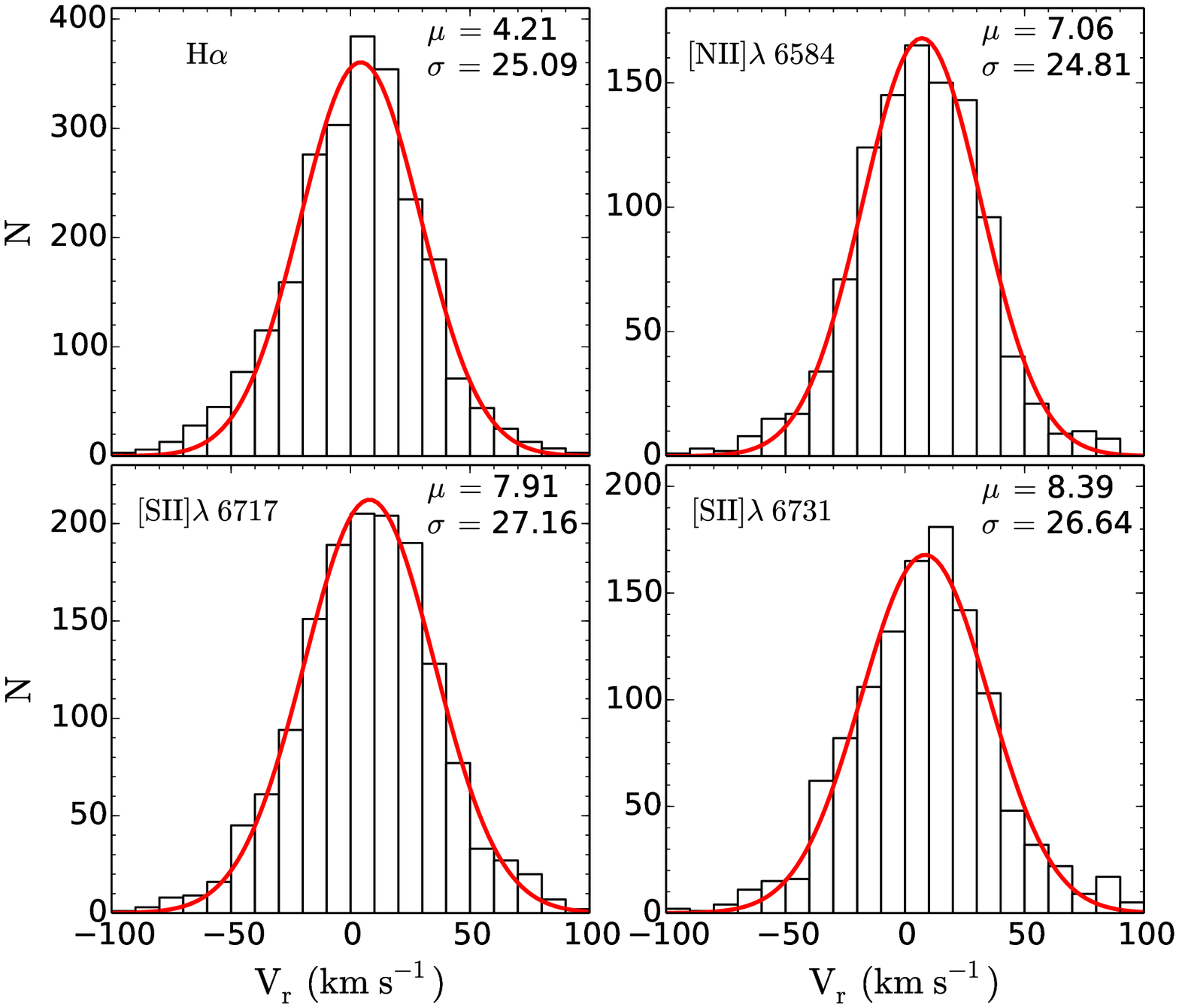}
    \caption{Histogram distributions of radial velocities determined from \Ha , \Lin{N}{II}{6584}, and \Lines{S}{II}{6717,\,6731} emission lines for spectra within 2$^\circ$ of the geometric centre of S147. Only radial velocity measurements of errors less than 20 \kms\ are included. Mean and standard deviation of the distribution are marked in each panel.}
    \label{fig:rv_hist}
\end{figure}

\section{Discussion}

\subsection{The radial velocity fields}

As shown in Fig.\,\ref{fig:spec-filament}, the \Lin{N}{II}{6548} line is in general rather weak and seen only in a few high quality spectra. \Ha\ is the strongest nebular line and visible in almost all the spectra. The \Lin{N}{II}{6584} line has a typical strength comparable to the \Lines{S}{II}{6717,\,6731} lines. The latter three lines have comparable occurrence rates, and are often less stronger than \Ha\ . For spectra of comparable SN ratio, the stronger the nebular lines are, the higher the detection rates will be, and vice verse. In the current work, only \Ha , \Lin{N}{II}{6584}, and \Lines{S}{II}{6717,\,6731} lines are used to measure the kinematic properties of S147. 

Fig.\,\ref{fig:rv-field} shows the radial velocity field of S147 as revealed by the emission lines measured. One can see that radial velocity fields mapped by emission lines exhibit spatial features well correlated. The radial velocities of different filaments of S147 show complex structure and variations. Some of the filaments are moving towards from us (such as the segment marked by blue ellipse in Fig.\,\ref{fig:rv-field}), some are moving away from us (such as the segment marked by red ellipse in Fig.\,\ref{fig:rv-field}). Probably due to the projection effects or a direct consequence of the complex structure of the filaments, many segments of filaments of blue- and red-shifted velocities are tangled together. One of such examples is marked by green ellipse in Fig.\,\ref{fig:rv-field}. The radial velocity fields presented here cover the whole extent of S147 with unprecedented sampling. Fig.\,\ref{fig:rv_hist} shows histogram distributions of all derived radial velocities of uncertainties less than 20\,\kms\ determined from various lines for spectra within 2$^\circ$ of the geometric centre of S147. The distributions peak between $\sim$\,0\,--\,10\,\kms .

\begin{figure}
    \centering
	\includegraphics[angle=270,width=85mm]{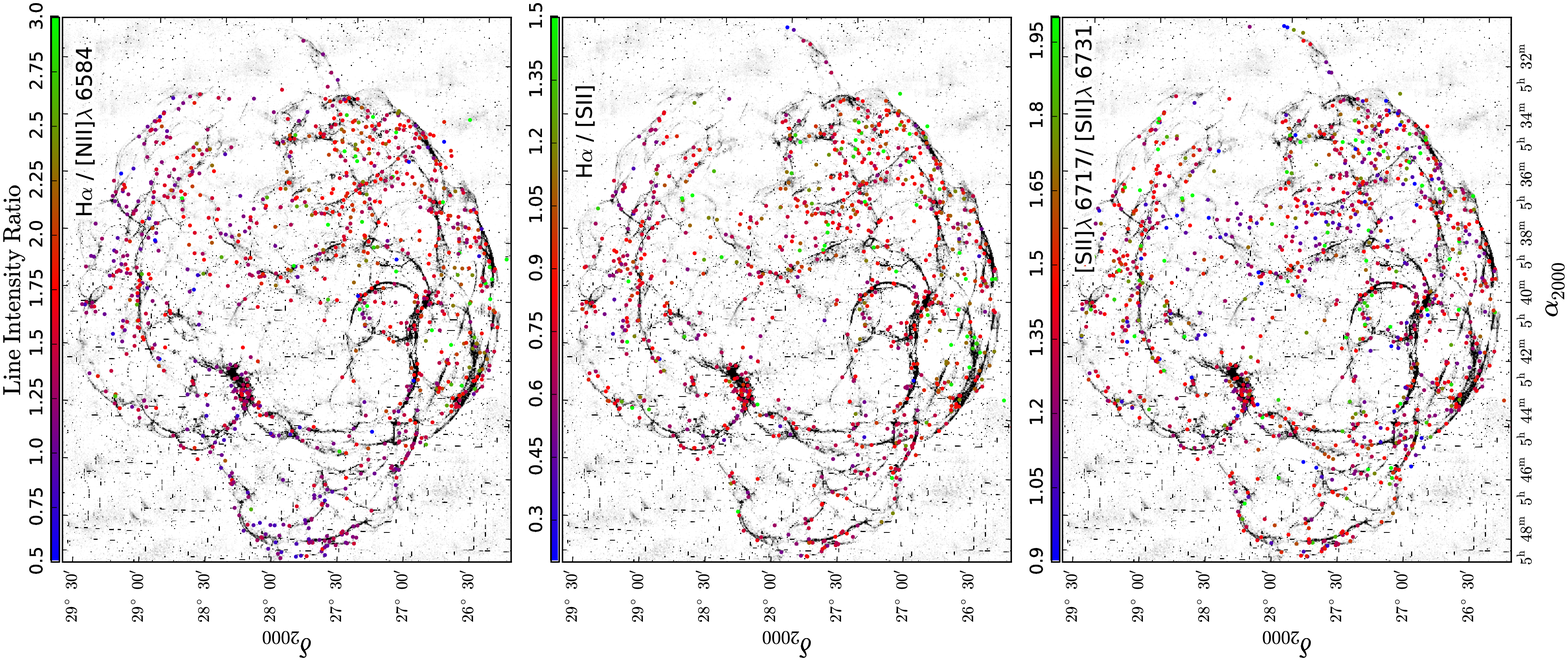}
    \caption{Similar to Fig.\,\ref{fig:rv-field}, but for line intensity ratios: \Ha /\Lin{N}{II}{6584}, \Ha /\Lines{S}{II}{6717,\,6731} and \Lin{S}{II}{6717}/\Lin{S}{II}{6731} (from top to bottom). Totally, within 3$^\circ$ of S147, there are 1075, 896, and 1041 data points in the panels from top to bottom, respectively.}
    \label{fig:line-intensity-ratios}
\end{figure}

\begin{figure}
    \centering
	\includegraphics[angle=0,width=170mm]{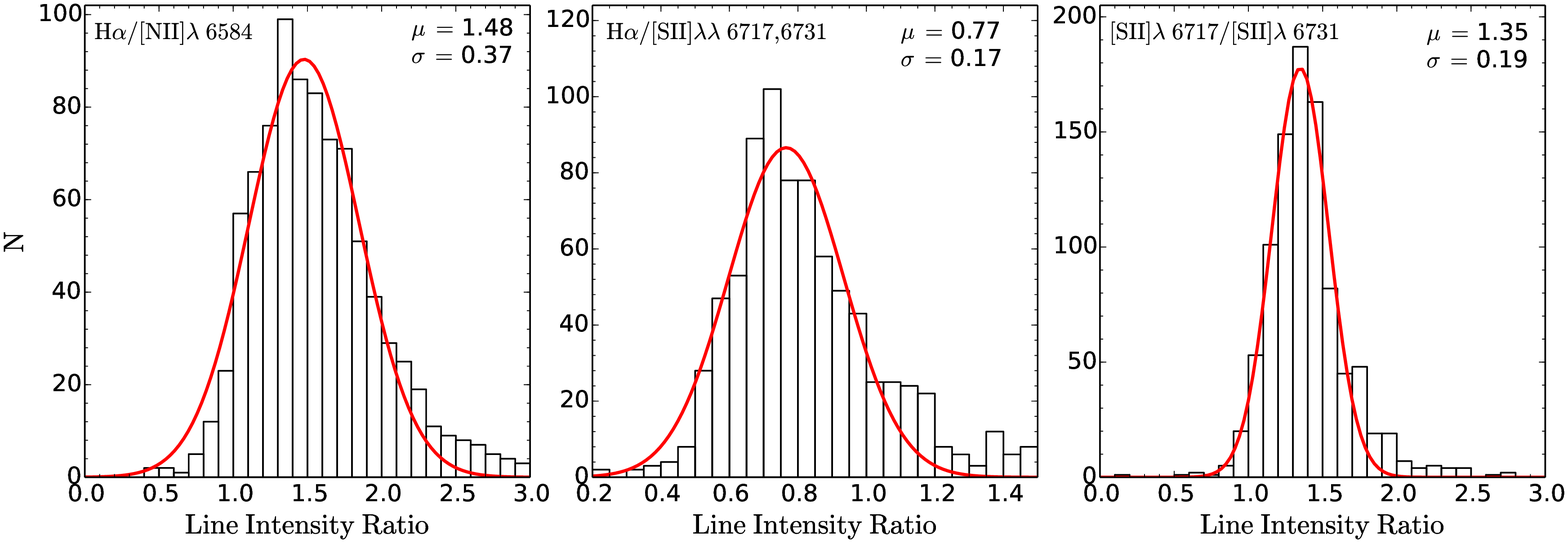}
    \caption{Histogram distributions of line intensity ratios \Ha /\Lin{N}{II}{6584}, \Ha /\Lines{S}{II}{6717,\,6731} and \Lin{S}{II}{6717}/\Lin{S}{II}{6731}, derived from spectra within 2$^\circ$ of the centre of S147. Red curves are Gaussian fits to the distributions, with mean $\mu$ and standard deviation $\sigma$ marked in each panel.}
    \label{fig:hist_line-intensity-ratio}
\end{figure}

\begin{figure}
    \centering
	\includegraphics[angle=0,width=85mm]{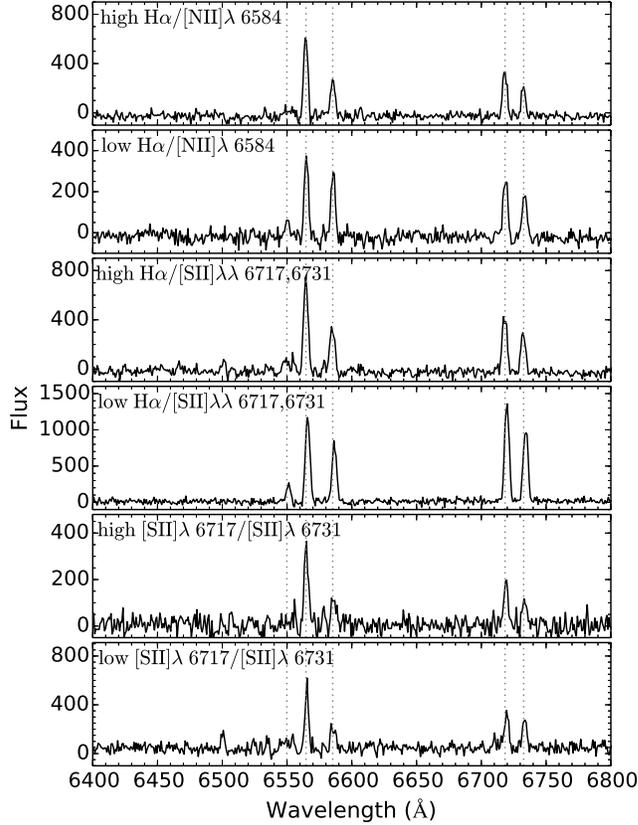}
    \caption{Example spectra of filaments with high and low line intensity ratios.}
    \label{fig:spec-ratio}
\end{figure}

\begin{figure}
    \centering
	\includegraphics[angle=0,width=85mm]{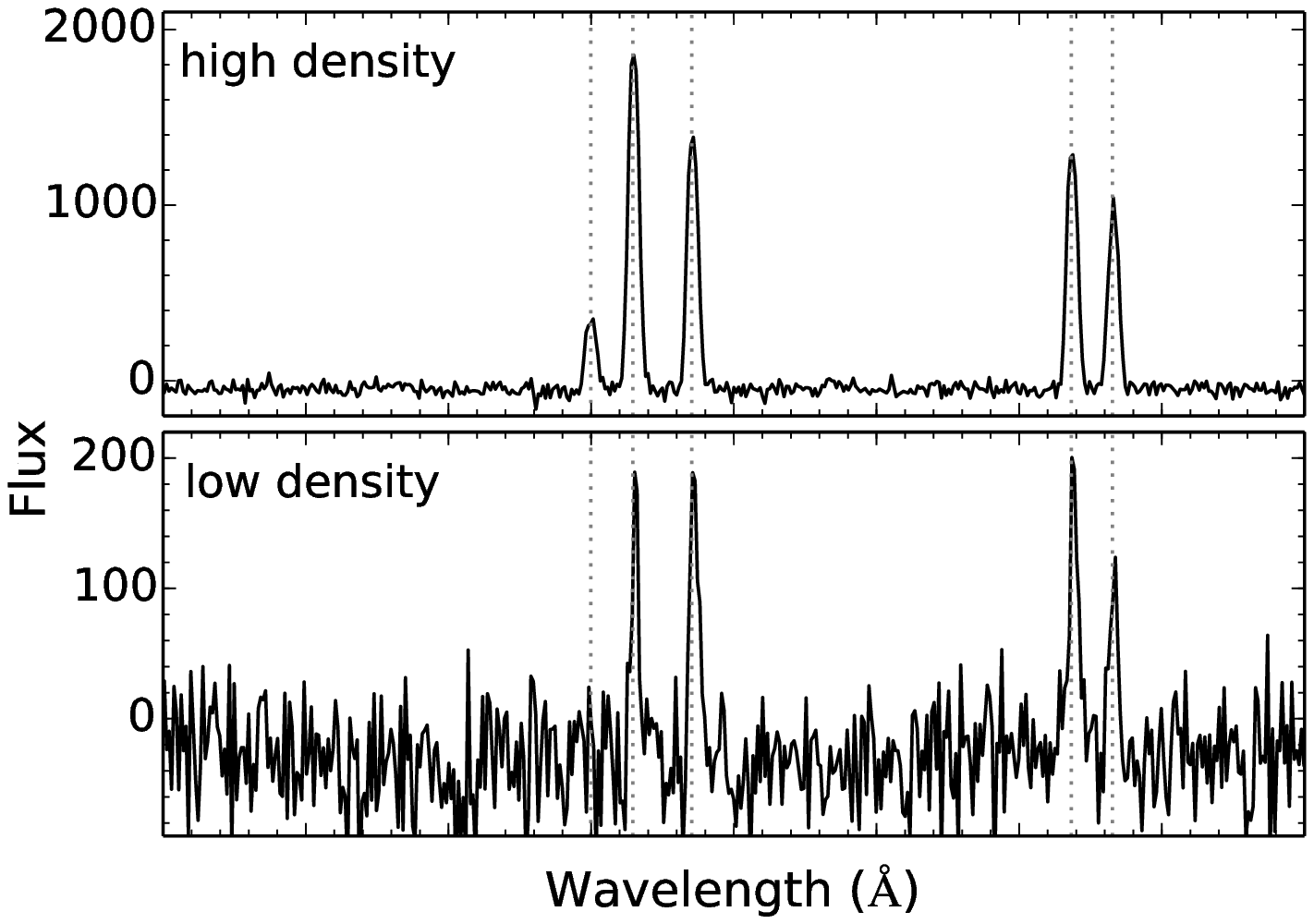}
    \caption{Example spectra of filaments with strong and weak line intensities.}
    \label{fig:spec-intensity}
\end{figure}

\subsection{Maps of line intensity ratios}

Fig.\,\ref{fig:line-intensity-ratios} shows the distributions of line intensity ratio \Ha/\Lin{N}{II}{6584}, \Ha/\Lines{S}{II}{6717,\,6731} and \Lin{S}{II}{6717}/\Lin{S}{II}{6731}. Similar to the distributions of radial velocities (Fig.\,\ref{fig:rv-field}), the line intensity ratios are well measured along prominent \Ha\ filaments. Fig.\,\ref{fig:hist_line-intensity-ratio} shows the histogram distributions of the above line intensity ratios, derived from spectra within 2$^\circ$ of the centre of S147. The ratios \Ha/\Lines{S}{II}{6717,\,6731} and \Lin{S}{II}{6717}/\Lin{S}{II}{6731} peak at $\sim$\,0.77 (dispersion $\sigma$\,=\,0.17) and 1.35 ($\sigma$\,=\,0.19; i.e. close to the low electron density limit), respectively. Our results are consistent with previous measurements (\Ha/\Lines{S}{II}{6717,\,6731} ranges between 0.7\,--\,1.08; \Lin{S}{II}{6717}/\Lin{S}{II}{6731} $\sim$\,1.4) in the literature \citep{Fesen1985ApJ...292...29F,Lozinskaia1976AZh....53...38L}. Since the \Lin{N}{II}{6548} line is weak and only detected in a very limited sample of spectra, we use only the \Lin{N}{II}{6584} line to calculate the \Ha/\Li{N}{II} ratio, i.e. \Ha/\Lin{N}{II}{6584}. The peak value of \Ha/\Lin{N}{II}{6584} is $\sim$\,1.48 ($\sigma$\,=\,0.37), again in good agreement with the literature values of $\sim$\,1.20\,--\,1.58 \citep{DOdorico1977A&AS...28..439D}. Figs.\ref{fig:spec-ratio} and \ref{fig:spec-intensity} show example spectra of filaments with high and low line intensity ratios and of strong and weak line intensities, respectively.

Generally SNRs differ from \Ion{H}{II} regions by showing stronger \Lines{S}{II}{6717,\,6731} emission. Quantitatively, \Ha/\Lines{S}{II}{6717,\,6731}\,$<$\,2.0 for SNRs \citep{Fesen1985ApJ...292...29F}. The middle panels of Fig.\,\ref{fig:line-intensity-ratios} and Fig.\,\ref{fig:hist_line-intensity-ratio} show that in S147, the \Ha/\Lines{S}{II}{6717,\,6731} ratio has values less than $\sim$\,1.4, consistent with S147 being a typical SNR.

It has been suggested that the intensity ratios \Ha/\Lines{N}{II}{6548,6584}, \Ha/\Lines{S}{II}{6717,\,6731} and \Lin{S}{II}{6717}/\Lin{S}{II}{6731} vary less from region to region of a given SNR than they do amongst different SNRs \citep{Daltabuit1976A&A....52...93D}. Based on this, \cite{Daltabuit1976A&A....52...93D} develop an evolutionary scheme of SNRs in which \Ha/\Lines{N}{II}{6548,6584} and \Lin{S}{II}{6717}/\Lin{S}{II}{6731} vary systematically as a function of the SNR diameter. In addition, the \Ha/\Lines{N}{II}{6548,6584} ratio combined with electron density indicated by the \Lin{S}{II}{6717}/\Lin{S}{II}{6731} ratio has been used to infer the elemental abundance gradient in the Milky Way, M31, and M33 \citep{Fesen1985ApJ...292...29F,Binette1982A&A...115..315B, Dopita1980ApJ...236..628D, Blair1981ApJ...247..879B, Blair1982ApJ...254...50B,Blair1985ApJ...289..582B}. This approach is only valid if the aforementioned line ratio does not vary significantly in a given SNR. Only a very limited filaments of Galactic SNRs have been studied spectroscopically. The extensive data presented in the current work provide an opportunity to investigate the intrinsic variations of line intensity ratios  in a given SNR and test the legitimacy of the aforementioned approach. 

From Fig.\,\ref{fig:line-intensity-ratios} and \ref{fig:hist_line-intensity-ratio}, we estimate that the intrinsic dispersions of line ratios amount to be approximately $\pm$14\% for \Lin{S}{II}{6717}/\Lin{S}{II}{6731}, $\pm$22\% for \Ha/\Lines{S}{II}{6717,\,6731}, and $\pm$25\% for \Ha/\Lin{N}{II}{6584}. The dispersions are larger than those previously estimated by \cite{Fesen1985ApJ...292...29F} based on only five observations (the mean values of \Ha/\Lines{N}{II}{6548,6584}, \Ha/\Lines{S}{II}{6717,\,6731} and \Lin{S}{II}{6717}/\Lin{S}{II}{6731} are respectively 1.28, 1.00, 1.61, with corresponding dispersions of $\pm$11\%, $\pm$8\%, and $\pm$12\%) and those estimated by \cite{Dodorico1976A&A....53..443D} based on only six observations (the mean values of \Ha/\Lines{N}{II}{6548,6584}, \Ha/\Lines{S}{II}{6717,\,6731} and \Lin{S}{II}{6717}/\Lin{S}{II}{6731} are respectively 1.44, 0.79, 1.35, with corresponding dispersions of $\pm$9\%, $\pm$10\%, and $\pm$6\%). On the other hand, the results presented here are comparable to the dispersions estimated for 9 typical SNRs as listed in Table 5 of \cite{Fesen1985ApJ...292...29F}. The large intrinsic dispersions of line intensity ratios seen in S147 cast doubt on the viability of using those ratios to infer the evolutionary stage of a SNR or to estimate the Galactic elemental abundance gradient using SNRs \citep[see also][]{Fesen1985ApJ...292...29F}.

Being an electron density diagnostic, the \Lin{S}{II}{6717}/\Lin{S}{II}{6731} ratio has a mean value of 1.35 for S147, i.e. close to the low density limit of 1.4 (see Fig 6. in \citet{Blair1985ApJ...289..582B} for the \Lin{S}{II}{6717}/\Lin{S}{II}{6731} ratio as a function of electron density, the electron density goes to the lowest when the \Lin{S}{II}{6717}/\Lin{S}{II}{6731} ratio has a value of 1.4). The result is consistent with typical values found for evolved SNRs \citep[larger than 1.10, thus implies an electron density lower than 300\,cm$^{-3}$;][]{Fesen1985ApJ...292...29F}. We estimated that \Lin{S}{II}{6717}/\Lin{S}{II}{6731} has an intrinsic dispersion of 14\% in S147, slightly larger than those found previously by \cite{Fesen1985ApJ...292...29F} and \cite{Dodorico1976A&A....53..443D} (12\% and 6\%, respectively). The result suggests a small range of electron densities and cloud pressures in S147. A relation of electron density and the \Lin{S}{II}{6717}\,/\,\Lin{S}{II}{6731} line ratio has previously been presented by \cite{Blair1985ApJ...289..582B} \citep[see][Fig.\,6]{Blair1985ApJ...289..582B} assuming a five level atomic model, the collisional strenghths of \cite{Pradhan1978MNRAS.183P..89P} and the transition probabilities of \cite{Mendoza1982MNRAS.198..127M}. From the plot of \cite{Blair1985ApJ...289..582B}, the mean value of \Lin{S}{II}{6717}/\Lin{S}{II}{6731} ratio seen in S147, 1.35 along with a standard deviation 0.19, implies an the electron density of only $\sim$\,50\,cm$^{-3}$, with an upper limit of about 200\,cm$^{-3}$.

\subsection{Comparison with previous spectroscopic observations}

\citet{Kirshner1979ApJ...229..147K} observed 15 positions of S147 (see Table 1 in their paper) and measured their radial velocities. Their Fig. 1 shows the slit positions used. The exact coordinates of the slit positions are however not given in the paper. So it is hard to make an exact comparison. Referring to their Fig. 1, we select some of their slit positions that overlap approximately with (or close to) some of the fiber positions measured with LAMOST. Table \ref{tab:comp_rv} compares the radial velocities determined by \citet{Kirshner1979ApJ...229..147K} and by us. Although the differences between our measurements and literature values for position `c' and `k' seem large ($\sim$\,30\,\kms), we need to remember that systematic uncertainties of radial velocity is 10\,\kms\ \citep{Luo2015RAA....15.1095L} and the comparison here is not exact for the spectra obtained in same position. So finally we conclude that the agreement is reasonable.

\begin{table}
	\centering
	\caption{The radial velocity comparison with those from \citet{Kirshner1979ApJ...229..147K}.}
	\label{tab:comp_rv}
	\begin{tabular}{ccc} 
		\hline
		Position & Radial Velocity & Radial Velocity  \\
		 & \citep{Kirshner1979ApJ...229..147K} & (here) \\
		         &  (\kms ) & (\kms )\\
		\hline
		c &	$+$54$\pm$7 & $+$21$\pm$8 \\
		d & $+$4$\pm$8 & $+$1$\pm$18 \\
		e & $-$21$\pm$9 & $-$20$\pm$6 \\
		h & $-$47$\pm$8 & $-$37$\pm$8 \\
		k & $+$33$\pm$6 & $+$66$\pm$7 \\
		m & $-$12$\pm$8 & $-$20$\pm$5 \\
		\hline
	\end{tabular}
\end{table}

\begin{table}
	\centering
	\caption{Comparison of line intensity ratio measurements with those of \citet{Fesen1985ApJ...292...29F}. The second column shows the distances between the positions of measurements of \citet{Fesen1985ApJ...292...29F} and of LAMOST.}
	\label{tab:comp_ratio}
	\begin{tabular}{cccc} 
		\hline
		 Position & Distance & \Ha/\Li{S}{II} & \Ha/\Li{S}{II} \\
		  & (arcmin) & \citep{Fesen1985ApJ...292...29F} & (here) \\
		  \hline
		P2 &	 3.5 & 0.90 & 0.87 \\
		P5 & 4.0 & 0.93 & 0.68 \\
		\hline
	\end{tabular}
\end{table}

\begin{figure}
    \centering
	\includegraphics[angle=270,width=85mm]{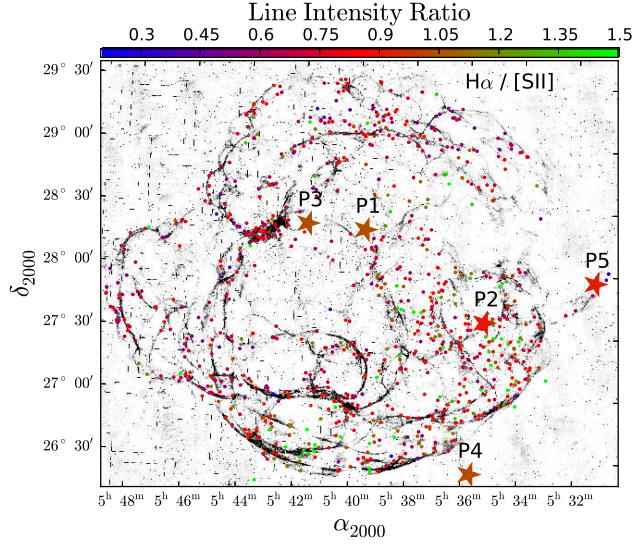}
    \caption{The five observations (the asterisk) of \citet{Fesen1985ApJ...292...29F} in our line intensity ratio map.}
    \label{fig:comp-fesen}
\end{figure}

\citet{Fesen1985ApJ...292...29F} carried out spectroscopic observations of five positions of S147. Two  of them are close to the filaments also observed with LAMOST (P2 and P5 in Fig.\ref{fig:comp-fesen}). Table \ref{tab:comp_ratio} compares the \Ha/\Li{S}{II} line intensity ratios as determined by \citet{Fesen1985ApJ...292...29F} and by us for P2 and P5. Because the exact positions of their observations are about 4 arcmin away from our observations, the small differences between the two sets of measurements seem to be acceptable.

\section{Summary}

We have analyzed all spectra available in the LAMOST DR2 collected in the vicinity field of SNR S147. The spectra are carefully sky subtracted. Both the spectra of sky background and of foreground and background stars are used. For high-quality stellar spectra with atmospheric parameters determined, the underlying stellar features are subtracted before measuring the nebular emission line properties. For stellar spectra of low SN ratio and with no stellar atmospheric parameters determined, the spectra are directly used to measure the nebular line properties. 

By measuring the prominent emission lines \Ha, \Lin{N}{II}{6584}, and \Lines{S}{II}{6717,\,6731}, we obtain fundamental kinematic properties of S147, including maps of radial velocity and intensity ratio of exquisite detail. As it is not a dedicated filamentary survey of LAMOST in S147 region, the number of filamentary spectra is very low comparing with the total spectral number in this region. But still this is first such as a large SNR has been fully mapped spectroscopically. 

The radial velocity distribution peaks $\sim$\,0\,--\,10\,\kms. The intensity ratios \Ha/\Lines{S}{II}{6717,\,6731} and \Lin{S}{II}{6717}/\Lin{S}{II}{6731} peak respectively at $\sim$\,0.77 ($\sigma$\,=\,0.17) and 1.35 ($\sigma$\,=\,0.19), consistent with previous determinations in the literature. The \Ha/\Lin{N}{II}{6584} ratios peak at $\sim$\,1.48 ($\sigma$\,=\,0.37), and agrees well with the literature value ($\sim$\,1.20\,--\,1.58). The intrinsic dispersions of line intensity ratios estimated here are larger than previous estimates in the literature, based on observations of only a few filaments. 

\normalem
\begin{acknowledgements}
This work is supported by Joint Funds of the National Natural Science Foundation of China (Grant No. U1531244) and the National Key Basic Research Program of China 2014CB845700. JJR acknowledges support from the Young Researcher  Grant of  National Astronomical Observatories, Chinese Academy of Sciences. JJR thanks Albert Zijlstra, Nick Wright, and Janet Drew for kindly providing the IPHAS \Ha\ image of S147. The LAMOST FELLOWSHIP is supported by Special Funding for Advanced Users, budgeted and administrated by Centre for Astronomical Mega-Science, Chinese Academy of Sciences (CAMS).

This work has made use of data products from the Guoshoujing Telescope (the Large Sky Area Multi-Object Fibre Spectroscopic Telescope, LAMOST). LAMOST is a National Major Scientific Project built by the Chinese Academy of Sciences. Funding for the project has been provided by the National Development and Reform Commission. LAMOST is operated and managed by the National Astronomical Observatories, Chinese Academy of Sciences.
\end{acknowledgements}
  
\bibliographystyle{raa}
\bibliography{bibtex}

\begin{thebibliography}{46}
\providecommand{\natexlab}[1]{#1}
\providecommand{\selectlanguage}[1]{\relax}

\bibitem[{{Anderson} et~al.(1996){Anderson}, {Cadwell}, {Jacoby}
  et~al.}]{Anderson1996ApJ...468L..55A}
{Anderson}, S.~B., {Cadwell}, B.~J., {Jacoby}, B.~A., et~al. 1996, \apjl, 468,
  L55

\bibitem[{{Binette} et~al.(1982){Binette}, {Dopita}, {Dodorico}, \&
  {Benvenuti}}]{Binette1982A&A...115..315B}
{Binette}, L., {Dopita}, M.~A., {Dodorico}, S., \& {Benvenuti}, P. 1982, \aap,
  115, 315

\bibitem[{{Blair} \& {Kirshner}(1985)}]{Blair1985ApJ...289..582B}
{Blair}, W.~P., \& {Kirshner}, R.~P. 1985, \apj, 289, 582

\bibitem[{{Blair} et~al.(1981){Blair}, {Kirshner}, \&
  {Chevalier}}]{Blair1981ApJ...247..879B}
{Blair}, W.~P., {Kirshner}, R.~P., \& {Chevalier}, R.~A. 1981, \apj, 247, 879

\bibitem[{{Blair} et~al.(1982){Blair}, {Kirshner}, \&
  {Chevalier}}]{Blair1982ApJ...254...50B}
{Blair}, W.~P., {Kirshner}, R.~P., \& {Chevalier}, R.~A. 1982, \apj, 254, 50

\bibitem[{{Chatterjee} et~al.(2009){Chatterjee}, {Brisken}, {Vlemmings}
  et~al.}]{Chatterjee2009ApJ...698..250C}
{Chatterjee}, S., {Brisken}, W.~F., {Vlemmings}, W.~H.~T., et~al. 2009, \apj,
  698, 250

\bibitem[{{Chen} et~al.(2017){Chen}, {Liu}, {Ren}
  et~al.}]{Chen2017MNRAS.472.3924C}
{Chen}, B.-Q., {Liu}, X.-W., {Ren}, J.-J., et~al. 2017, \mnras, 472, 3924

\bibitem[{{Clark} \& {Caswell}(1976)}]{Clark1976MNRAS.174..267C}
{Clark}, D.~H., \& {Caswell}, J.~L. 1976, \mnras, 174, 267

\bibitem[{{Cui} et~al.(2012){Cui}, {Zhao}, {Chu}
  et~al.}]{Cui2012RAA....12.1197C}
{Cui}, X.-Q., {Zhao}, Y.-H., {Chu}, Y.-Q., et~al. 2012, Research in Astronomy
  and Astrophysics, 12, 1197

\bibitem[{{Daltabuit} et~al.(1976){Daltabuit}, {Dodorico}, \&
  {Sabbadin}}]{Daltabuit1976A&A....52...93D}
{Daltabuit}, E., {Dodorico}, S., \& {Sabbadin}, F. 1976, \aap, 52, 93

\bibitem[{{Deng} et~al.(2012){Deng}, {Newberg}, {Liu}
  et~al.}]{Deng2012RAA....12..735D}
{Deng}, L.-C., {Newberg}, H.~J., {Liu}, C., et~al. 2012, Research in Astronomy
  and Astrophysics, 12, 735

\bibitem[{{Din{\c c}el} et~al.(2015){Din{\c c}el}, {Neuh{\"a}user}, {Yerli}
  et~al.}]{Dincel2015MNRAS.448.3196D}
{Din{\c c}el}, B., {Neuh{\"a}user}, R., {Yerli}, S.~K., et~al. 2015, \mnras,
  448, 3196

\bibitem[{{Dodorico} \& {Sabbadin}(1976)}]{Dodorico1976A&A....53..443D}
{Dodorico}, S., \& {Sabbadin}, F. 1976, \aap, 53, 443

\bibitem[{{D'Odorico} \& {Sabbadin}(1977)}]{DOdorico1977A&AS...28..439D}
{D'Odorico}, S., \& {Sabbadin}, F. 1977, \aaps, 28, 439

\bibitem[{{Dopita} et~al.(1980){Dopita}, {Dodorico}, \&
  {Benvenuti}}]{Dopita1980ApJ...236..628D}
{Dopita}, M.~A., {Dodorico}, S., \& {Benvenuti}, P. 1980, \apj, 236, 628

\bibitem[{{Drew} et~al.(2005){Drew}, {Greimel}, {Irwin}
  et~al.}]{Drew2005MNRAS.362..753D}
{Drew}, J.~E., {Greimel}, R., {Irwin}, M.~J., et~al. 2005, \mnras, 362, 753

\bibitem[{{Fesen} et~al.(1985){Fesen}, {Blair}, \&
  {Kirshner}}]{Fesen1985ApJ...292...29F}
{Fesen}, R.~A., {Blair}, W.~P., \& {Kirshner}, R.~P. 1985, \apj, 292, 29

\bibitem[{{Fitzpatrick}(1999)}]{Fitzpatrick1999PASP..111...63F}
{Fitzpatrick}, E.~L. 1999, \pasp, 111, 63

\bibitem[{{Fuerst} \& {Reich}(1986)}]{Fuerst1986A&A...163..185F}
{Fuerst}, E., \& {Reich}, W. 1986, \aap, 163, 185

\bibitem[{{Green}(2014)}]{Green2014BASI...42...47G}
{Green}, D.~A. 2014, Bulletin of the Astronomical Society of India, 42, 47

\bibitem[{{Guseinov} et~al.(2003){Guseinov}, {Ankay}, {Sezer}, \&
  {Tagieva}}]{Guseinov2003A&AT...22..273G}
{Guseinov}, O.~H., {Ankay}, A., {Sezer}, A., \& {Tagieva}, S.~O. 2003,
  Astronomical and Astrophysical Transactions, 22, 273

\bibitem[{{Katsuta} et~al.(2012){Katsuta}, {Uchiyama}, {Tanaka}
  et~al.}]{Katsuta2012ApJ...752..135K}
{Katsuta}, J., {Uchiyama}, Y., {Tanaka}, T., et~al. 2012, \apj, 752, 135

\bibitem[{{Kirshner} \& {Arnold}(1979)}]{Kirshner1979ApJ...229..147K}
{Kirshner}, R.~P., \& {Arnold}, C.~N. 1979, \apj, 229, 147

\bibitem[{{Kundu} et~al.(1980){Kundu}, {Angerhofer}, {Fuerst}, \&
  {Hirth}}]{Kundu1980A&A....92..225K}
{Kundu}, M.~R., {Angerhofer}, P.~E., {Fuerst}, E., \& {Hirth}, W. 1980, \aap,
  92, 225

\bibitem[{{Liu} et~al.(2014){Liu}, {Yuan}, {Huo}
  et~al.}]{Liu2014IAUS..298..310L}
{Liu}, X.-W., {Yuan}, H.-B., {Huo}, Z.-Y., et~al. 2014, in Setting the scene
  for Gaia and LAMOST, \emph{IAU Symposium}, vol. 298, edited by S.~{Feltzing},
  G.~{Zhao}, N.~A. {Walton}, \& P.~{Whitelock}, 310--321

\bibitem[{{Lozinskaia}(1976)}]{Lozinskaia1976AZh....53...38L}
{Lozinskaia}, T.~A. 1976, \azh, 53, 38

\bibitem[{{Luo} et~al.(2012){Luo}, {Zhang}, {Zhao}
  et~al.}]{Luo2012RAA....12.1243L}
{Luo}, A.-L., {Zhang}, H.-T., {Zhao}, Y.-H., et~al. 2012, Research in Astronomy
  and Astrophysics, 12, 1243

\bibitem[{{Luo} et~al.(2015){Luo}, {Zhao}, {Zhao}
  et~al.}]{Luo2015RAA....15.1095L}
{Luo}, A.-L., {Zhao}, Y.-H., {Zhao}, G., et~al. 2015, Research in Astronomy and
  Astrophysics, 15, 1095

\bibitem[{{Mendoza} \& {Zeippen}(1982)}]{Mendoza1982MNRAS.198..127M}
{Mendoza}, C., \& {Zeippen}, C.~J. 1982, \mnras, 198, 127

\bibitem[{{Minkowski}(1958)}]{Minkowski1958RvMP...30.1048M}
{Minkowski}, R. 1958, Reviews of Modern Physics, 30, 1048

\bibitem[{{Ng} et~al.(2007){Ng}, {Romani}, {Brisken}, {Chatterjee}, \&
  {Kramer}}]{Ng2007ApJ...654..487N}
{Ng}, C.-Y., {Romani}, R.~W., {Brisken}, W.~F., {Chatterjee}, S., \& {Kramer},
  M. 2007, \apj, 654, 487

\bibitem[{{Phillips} et~al.(1981){Phillips}, {Gondhalekar}, \&
  {Blades}}]{Phillips1981MNRAS.195..485P}
{Phillips}, A.~P., {Gondhalekar}, P.~M., \& {Blades}, J.~C. 1981, \mnras, 195,
  485

\bibitem[{{Pradhan}(1978)}]{Pradhan1978MNRAS.183P..89P}
{Pradhan}, A.~K. 1978, \mnras, 183, 89P

\bibitem[{{Sauvageot} et~al.(1990){Sauvageot}, {Ballet}, \&
  {Rothenflug}}]{Sauvageot1990A&A...227..183S}
{Sauvageot}, J.~L., {Ballet}, J., \& {Rothenflug}, R. 1990, \aap, 227, 183

\bibitem[{{Stoughton} et~al.(2002){Stoughton}, {Lupton}, {Bernardi}
  et~al.}]{Stoughton2002AJ....123..485S}
{Stoughton}, C., {Lupton}, R.~H., {Bernardi}, M., et~al. 2002, \aj, 123, 485

\bibitem[{{Sun} et~al.(1996){Sun}, {Anderson}, {Aschenbach}
  et~al.}]{Sun1996rftu.proc..195S}
{Sun}, X., {Anderson}, S., {Aschenbach}, B., et~al. 1996, in Roentgenstrahlung
  from the Universe, edited by H.~U. {Zimmermann}, J.~{Tr{\"u}mper}, \&
  H.~{Yorke}, 195--196

\bibitem[{{Voges} et~al.(1999){Voges}, {Aschenbach}, {Boller}
  et~al.}]{Voges1999A&A...349..389V}
{Voges}, W., {Aschenbach}, B., {Boller}, T., et~al. 1999, \aap, 349, 389

\bibitem[{{Woltjer}(1972)}]{Woltjer1972ARA&A..10..129W}
{Woltjer}, L. 1972, \araa, 10, 129

\bibitem[{{Xiang} et~al.(2015{\natexlab{a}}){Xiang}, {Liu}, {Yuan}
  et~al.}]{Xiang2015MNRAS.448...90X}
{Xiang}, M.~S., {Liu}, X.~W., {Yuan}, H.~B., et~al. 2015{\natexlab{a}}, \mnras,
  448, 90

\bibitem[{{Xiang} et~al.(2015{\natexlab{b}}){Xiang}, {Liu}, {Yuan}
  et~al.}]{Xiang2015MNRAS.448..822X}
{Xiang}, M.~S., {Liu}, X.~W., {Yuan}, H.~B., et~al. 2015{\natexlab{b}}, \mnras,
  448, 822

\bibitem[{{Xiang} et~al.(2017){Xiang}, {Liu}, {Yuan}
  et~al.}]{Xiang2017MNRAS.467.1890X}
{Xiang}, M.-S., {Liu}, X.-W., {Yuan}, H.-B., et~al. 2017, \mnras, 467, 1890

\bibitem[{{Xiao} et~al.(2008){Xiao}, {F{\"u}rst}, {Reich}, \&
  {Han}}]{Xiao2008A&A...482..783X}
{Xiao}, L., {F{\"u}rst}, E., {Reich}, W., \& {Han}, J.~L. 2008, \aap, 482, 783

\bibitem[{{Yuan} et~al.(2015){Yuan}, {Liu}, {Huo}
  et~al.}]{Yuan2015MNRAS.448..855Y}
{Yuan}, H.-B., {Liu}, X.-W., {Huo}, Z.-Y., et~al. 2015, \mnras, 448, 855

\bibitem[{{Yuan} et~al.(2013){Yuan}, {Liu}, \&
  {Xiang}}]{Yuan2013MNRAS.430.2188Y}
{Yuan}, H.~B., {Liu}, X.~W., \& {Xiang}, M.~S. 2013, \mnras, 430, 2188

\bibitem[{{Zhang} et~al.(2014){Zhang}, {Liu}, {Yuan}
  et~al.}]{Zhang2014RAA....14..456Z}
{Zhang}, H.-H., {Liu}, X.-W., {Yuan}, H.-B., et~al. 2014, Research in Astronomy
  and Astrophysics, 14, 456-470

\bibitem[{{Zhao} et~al.(2012){Zhao}, {Zhao}, {Chu}, {Jing}, \&
  {Deng}}]{Zhao2012RAA....12..723Z}
{Zhao}, G., {Zhao}, Y.-H., {Chu}, Y.-Q., {Jing}, Y.-P., \& {Deng}, L.-C. 2012,
  Research in Astronomy and Astrophysics, 12, 723

\end{thebibliography}

\end{document}